\documentclass[aps,prl,twocolumn,secnumarabic,balancelastpage,amsmath,amssymb,floatfix,superscriptaddress,longbibliography]{revtex4-1}
\usepackage{amssymb}
\usepackage{amsmath}
\usepackage{amsfonts}
\usepackage{braket}
\usepackage{graphicx}
\usepackage{epstopdf}
\usepackage{enumerate}
\usepackage{natbib}
\usepackage{epsfig}
\usepackage[toc,page,title,titletoc,header]{appendix}
\usepackage{dsfont,amsthm,amsbsy}
\usepackage{todonotes}
\usepackage{hyperref}
\usepackage{cancel}

\usepackage[normalem]{ulem}

\newcommand{\local}{\mathrm{local}}
\newcommand{\heat}{\mathrm{heat}}
\newcommand{\ave}{\mathrm{avg}}

\newcommand{\figOne}{
  \begin{figure}[t]
    \includegraphics[width=3.1in]{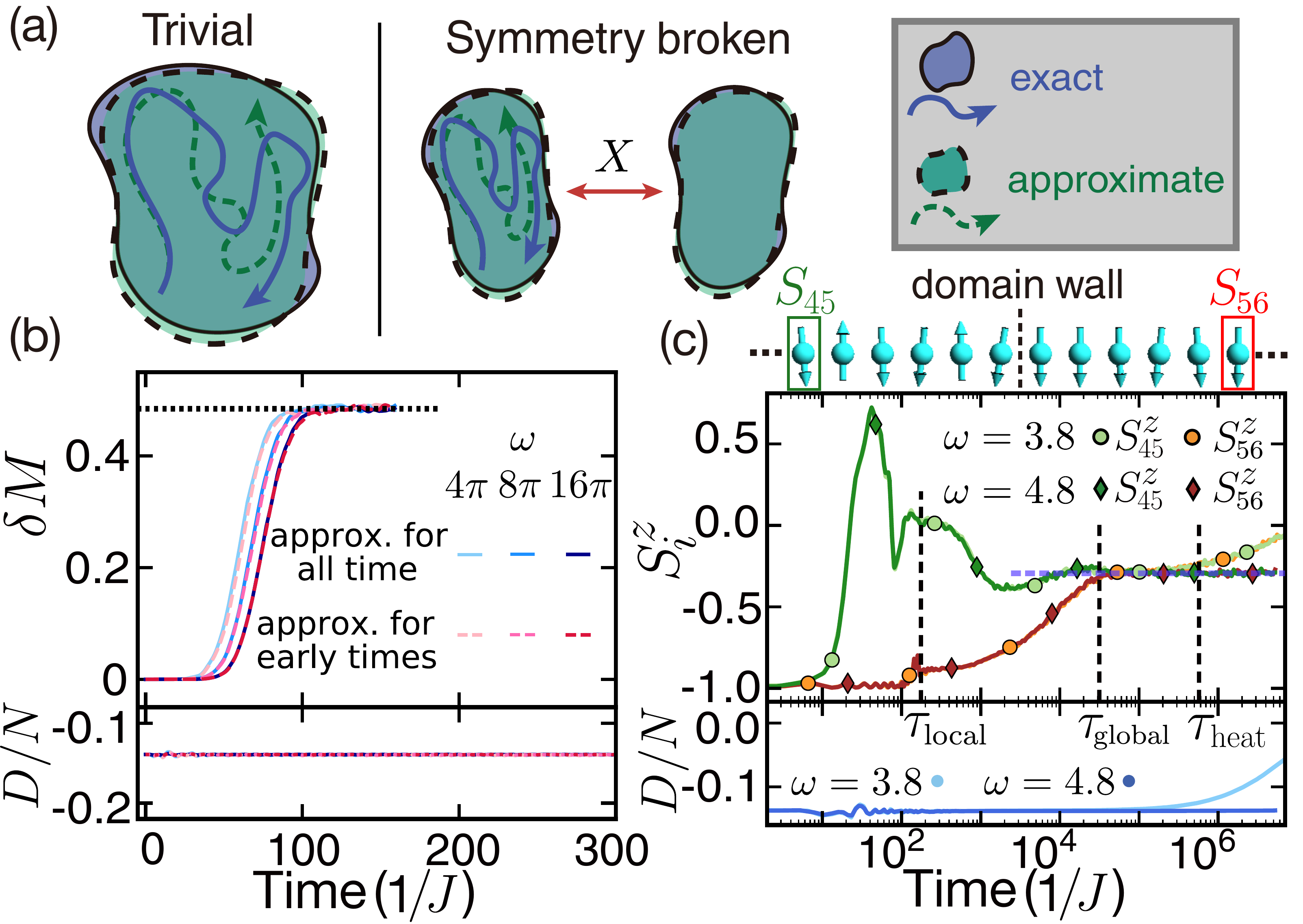}
    \centering
    \caption{
    (a) Schematic depicting trajectories in a classical phase space. The exact Floquet trajectory (blue) diverges from the approximate trajectory under the effective Hamiltonian (green).  However, the exact evolution of a \emph{finite region} in phase space is well-captured by the effective Hamiltonian.  
    (b) The  dynamics  of  the  magnetization difference, $\delta M(t)$, and the energy density, $D/N$, for a single initial state with $N=10^4$.
    Solid lines depict approximate evolution under $D$ for all times. 
    Dashed lines indicate approximate evolution under $D$ for short times ($t\le 1/J$), followed by exact Floquet evolution. 
    Agreement between solid and dashed curves highlights the role of classical chaos in the growth of errors. 
    While errors in local observables [i.e.~$\delta M(t)$] accumulate rapidly, the energy density remains conserved throughout the dynamics.
    (c) The prethermal dynamics of an ensemble of initial states quickly converge with increasing frequency.
    Before Floquet heating brings the system to infinite temperature, the magnetization approaches the value associated with the corresponding prethermal ensemble of $D$ (blue dashed line, computed via Monte Carlo \cite{SM}).
    }
    \label{fig1}
  \end{figure}
}

\newcommand{\figTwo}{
  \begin{figure}[t]
    \includegraphics[width=3.4in]{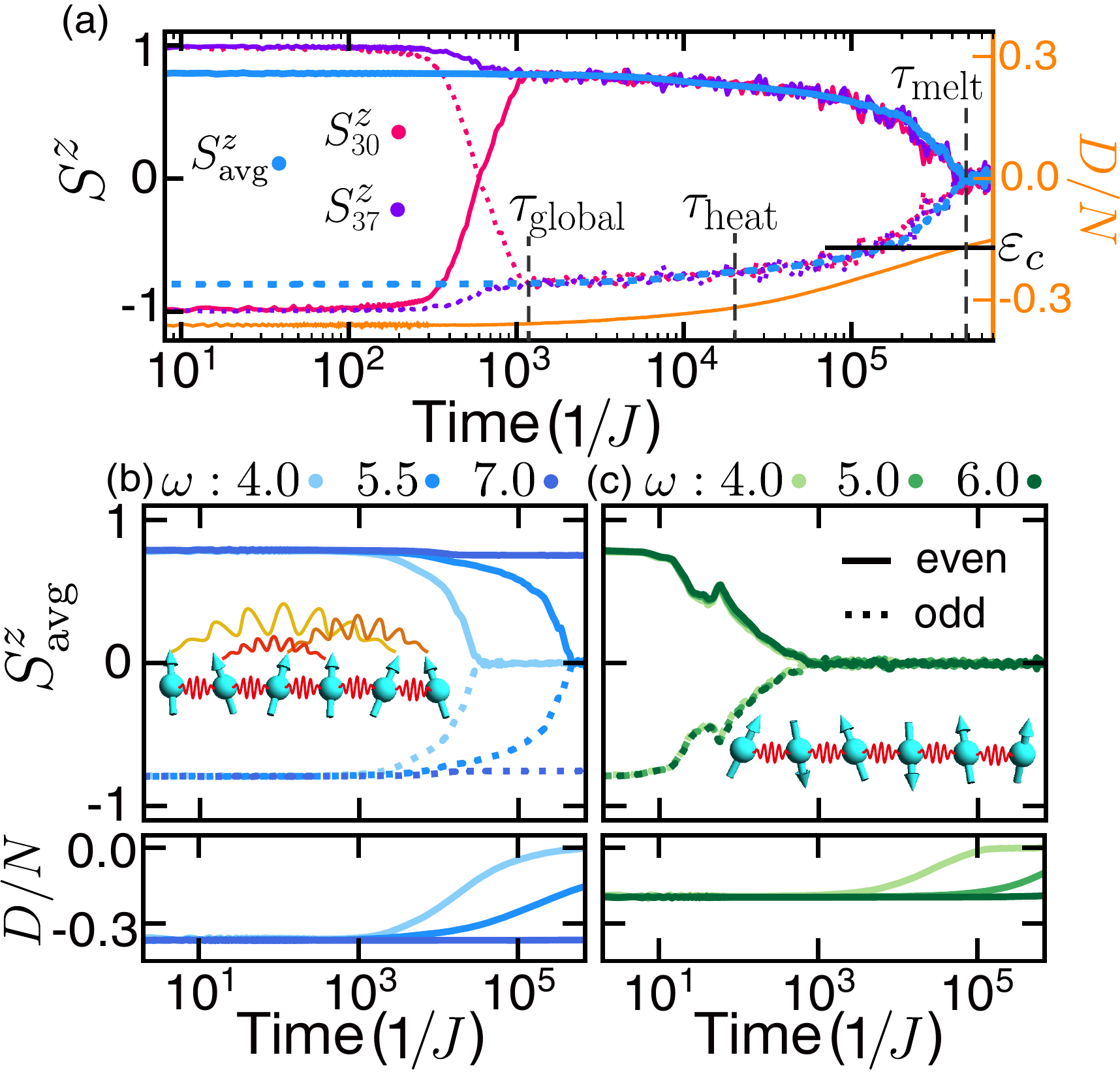}
    \centering
    \caption{
      (a) Dynamics of a classical prethermal time crystal in a one-dimensional long-range interacting spin chain.
      At $\tau_\mathrm{global}$, different sites exhibit the same magnetization, indicating  equilibration. 
      For an exponentially long intermediate time window, $\tau_\mathrm{global} < t< \tau_\mathrm{melt}$, the system oscillates between positive and negative magnetization values for even (solid line) and odd periods (dotted line). This subharmonic response remains stable until the energy density crosses $\varepsilon_c$ and the CPDTC melts. 
      (b,c) Prethermal dynamics of the spin chain for different frequencies $\omega$ with  either  long-range [b] or short-range [c] interactions.
      For long-range interactions, 
      the lifetime of the CPDTC is exponentially enhanced by increasing the frequency of the drive.
      For short-range interactions, transient period doubling decays at a frequency independent timescale, which is significantly shorter than the Floquet heating time (bottom panel). 
    }
    \label{fig2}
  \end{figure}
}

\newcommand{\figThree}{
  \begin{figure}[t]
    \includegraphics[width=3.2in]{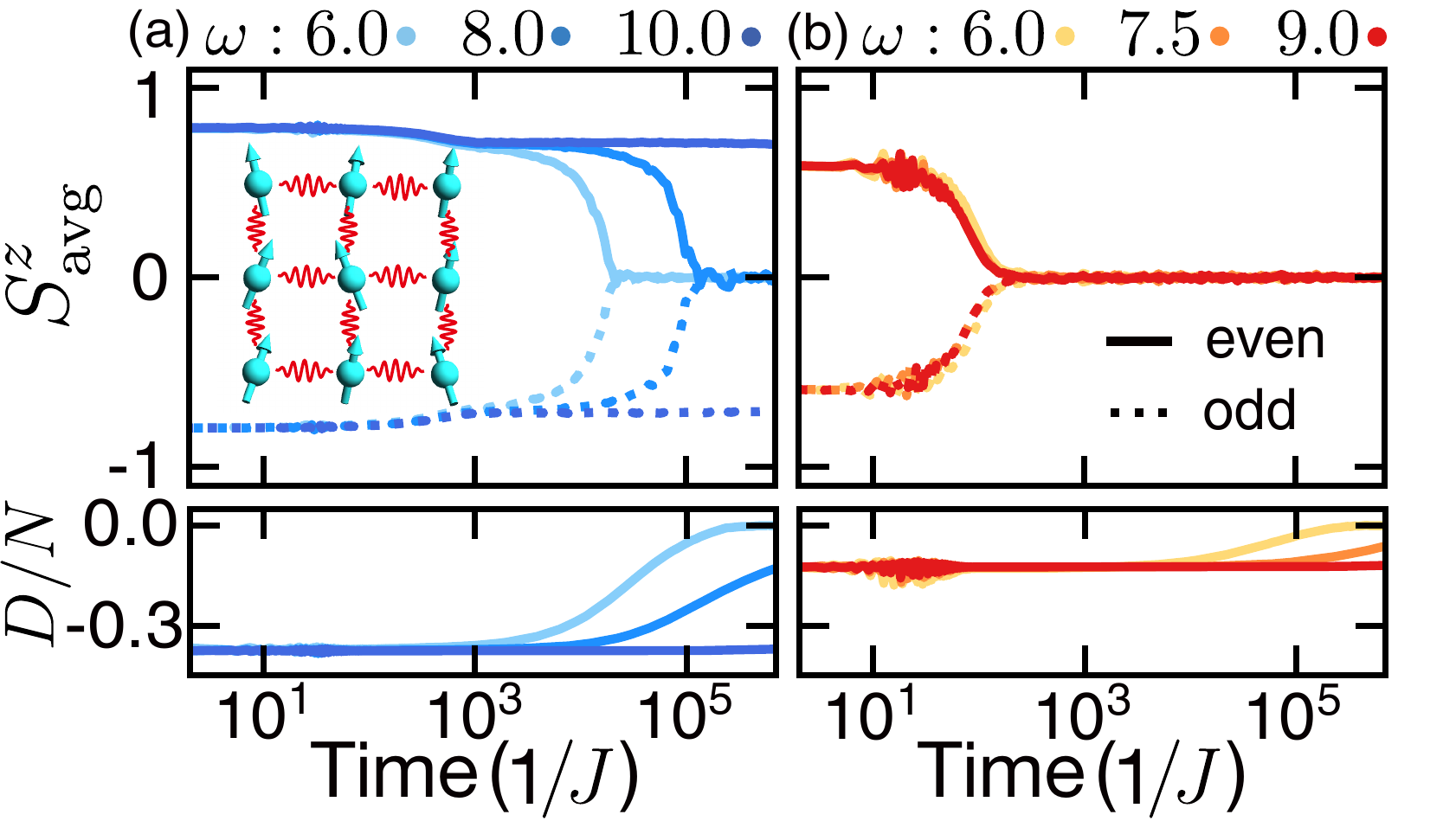}
    \centering
    \caption{
    Prethermal dynamics of a nearest-neighbor interacting classical spin model on the square lattice. (a) For a low-energy-density initial state, the system exhibits robust period doubling until  exponentially late times. (b) For a high-energy-density initial state, the magnetization decays to zero rapidly, well before the Floquet heating time.  
    This highlights the presence of a critical energy density and the importance of symmetry-breaking for the existence of a CPDTC. 
    }
    \label{fig3}
  \end{figure}
}

\begin{document}
\title{Floquet phases of matter via classical prethermalization}
\author{Bingtian Ye}
\affiliation{Department of Physics, University of California, Berkeley, CA 94720, USA}
\author{Francisco Machado}
\affiliation{Department of Physics, University of California, Berkeley, CA 94720, USA}
\affiliation{Materials Science Division, Lawrence Berkeley National Laboratory, Berkeley, CA 94720, USA}
\author{Norman Y. Yao}
\affiliation{Department of Physics, University of California, Berkeley, CA 94720, USA}
\affiliation{Materials Science Division, Lawrence Berkeley National Laboratory, Berkeley, CA 94720, USA}

\date{\today}
\begin{abstract}
We demonstrate that the prethermal regime of periodically-driven (Floquet), classical many-body systems can host non-equilibrium phases of matter. 
In particular, we show that there exists an effective Hamiltonian, which captures the dynamics of ensembles of classical trajectories, despite the breakdown of this description at the single trajectory level. 
In addition, we prove that the effective Hamiltonian can host emergent symmetries protected by the discrete time-translation symmetry of the drive. 
The spontaneous breaking of such an emergent symmetry leads to a sub-harmonic response, characteristic of time crystalline order, that survives to exponentially late times in the frequency of the drive. 
To this end, we numerically demonstrate the existence of classical prethermal time crystals in systems with different dimensionalities and ranges of interaction.
Extensions to higher order and fractional time crystals are also discussed. 
\end{abstract}


\maketitle

Many-body Floquet systems can host a variety of intrinsically  non-equilibrium phases of matter~\cite{Inoue10,Lindner11a,Jiang11b,else:2016,khemani:2016,yao:2017,potter2016classification,Potirniche_2017}. 
One of the central challenges in stabilizing such phases is the presence of Floquet heating---a generic interacting system will absorb energy from the driving field until it approaches a featureless, infinite temperature state~\cite{prosen1998time,prosen1999ergodic,lazarides2014equilibrium,bukov2015universal,d'alessio_drivenlongtime_2014}.
%
In quantum systems, strong disorder can induce many-body localization (MBL) which prevents  Floquet heating and enables the system to remain in a non-equilibrium steady state  until arbitrarily late times~\cite{lazarides_2014,d'alessio_drivenlongtime_2014,ponte2015periodically,abanin:2019}. 
%
%
Since localization relies upon the discreteness of energy levels, this specific approach is intrinsically quantum mechanical and naturally begs the following question: To what extent do Floquet non-equilibrium phases require either quantum mechanics or disorder \cite{lazarides2017fate,zhu:2019,dai2019truncated,alekseev2020provenance,yao2020classical,pizzi:2020, lazarides:2020, pizzi2021higher, pizzi:2021}?

An elegant, but partial, answer to this question is provided within the framework of Floquet prethermalization in disorder-free systems~\cite{berges2004prethermalization,abanin2015exponentially,machado_exponentially_2017,ho2018bounds,kuwahara2016floquet,mori2016rigorous,abanin2017rigorous,abanin2017effective,ye2020emergent,rajak2019characterizations,mori2018floquet,howell2019asymptotic,hodson:2021,kyprianidis:2021}. 
When the driving frequency, $\omega$, is larger than the system's local energy scale, $J_{\local}$, Floquet heating is suppressed until exponentially late times, $\tau_{\textrm{heat}} \sim e^{\omega / J_{\local}}$.
In particular, directly absorbing energy from the drive is highly off-resonant, and heating only occurs via higher order processes that involve multiple, correlated local rearrangements. %
This simple physical intuition holds for both quantum and classical systems. 

In the quantum setting, Floquet prethermalization has an additional feature: There exists an effective Hamiltonian that accurately captures the dynamics of the system until $\tau_{\textrm{heat}}$. 
Whenever the periodic drive induces an emergent symmetry in this effective Hamiltonian, novel non-equilibrium prethermal phases of matter, such as discrete time crystals or Floquet symmetry-protected topological phases, can emerge  \cite{Inoue10,Lindner11a,Jiang11b,else:2016,khemani:2016,yao:2017,Potirniche_2017,else2017prethermal,machado2020long,kyprianidis:2021,potter2016classification,else:2016b,vonkeyserlingk:2016,roy:2016,roy:2017,rudner:2019,oka:2019}.
%
Whether analogous phases are also possible in
classical many-body systems is significantly more subtle; in particular, although classical prethermalization features slow Floquet heating, there is no effective Hamiltonian that accurately captures the prethermal dynamics~\cite{mori2018floquet}.

\figOne

In this Letter, we show that the lack of an effective Hamiltonian does not preclude the existence of novel, non-equilibrium phases in classical Floquet systems; we highlight this by explicitly constructing a classical prethermal discrete time crystal (CPDTC). 
Our main results are three fold. 
First, we demonstrate that the inability of an effective Hamiltonian to generate the Floquet dynamics is a direct consequence of classical chaos---small errors at early times lead to exponentially diverging single trajectories. 
This connection to chaos suggests that one should 
forgo the focus on individual trajectories and rather ask whether there is an effective Hamiltonian that captures the prethermal dynamics of an \emph{ensemble} of trajectories (Fig.~\ref{fig1}).
We show that this is indeed the case. 
Second, we prove that, much like the quantum case, the effective Hamiltonian can host an emergent symmetry which is protected by the discrete time translation symmetry of the periodic drive. 
%
%
Finally, we propose, analyze and numerically simulate a variety of different classical prethermal time crystals in one and two dimensions.
%



\emph{Prethermalization in classical dynamics}---Consider a classical Floquet Hamiltonian, $H_F(t)=H_F(t+T)$, with period $T = 2\pi / \omega$.
%
For $\omega \gg J_{\local}$, one can construct a perturbative expansion of the Floquet dynamics in powers of $J_{\local} / \omega$ \cite{ft1}.
In general, this Floquet-Magnus expansion diverges, reflecting the many-body system's 
 late-time approach to infinite temperature (via energy absorption from the drive). 
However, when truncated at an appropriate order, $n^* \sim \omega/J_{\local}$, the expansion defines a static Hamiltonian, $D$, which remains quasi-conserved for exponentially long times (under the full Floquet dynamics)  \cite{abanin2015exponentially,ho2018bounds,mori2018floquet}:
\begin{equation} 
\frac{1}{N}|D(t=mT)-D(t=0)|<mJ_{\local}\cdot\mathcal{O}(e^{-\omega/J_{\local}}),
\label{Eq:energy}
\end{equation}
where $N$ is the system size and $m \in \mathbb{N}$ is the number of  Floquet cycles.
%
To this end, Eqn.~\ref{Eq:energy} precisely formalizes the existence of an intermediate, prethermal regime.
In particular, for times 
$t<\tau_\mathrm{heat}\sim \mathcal{O}(e^{\omega/J_{\local}})$, the energy density of the system (measured with respect to $D$), remains  constant up to $\sim \mathcal{O}(J_{\local})$.
%
%

Nevertheless, the question remains: Is $D$ also the effective prethermal Hamiltonian, which generates the dynamics before $\tau_\textrm{heat}$? 
In the quantum setting, the answer is yes \cite{ft2,mori2016rigorous, abanin2017rigorous, machado2020long}. 
However, in classical systems, $D$ is only proven to faithfully reproduce the Floquet evolution over a \emph{single} driving period \cite{mori2018floquet}:
\begin{equation}
|O(T)-O'(T)| \le \mathcal{O}(e^{-\omega/J_{local}}).
\label{Eq:ob_c}
\end{equation}
Here, $O$ is a generic local observable and $O(T)$ represents its evolution under the full Floquet Hamiltonian [i.e.~$H_F(t)$], while $O'(T)$ represents its evolution under $D$. 
Note that hereon out, observables with a prime, will always correspond to evolution under $D$. 

Naively, one might expect the single period errors in Eqn.~2 to accumulate additively as one evolves to later times.
%
However, this does not account for compounding effects, where early-time errors propagate through the many-body system and induce additional deviations. 
In the quantum case, the existence of Lieb-Robinson bounds constrains the propagation of errors and enables one to prove that deviations grow algebraically in the number of Floquet cycles:  $|O(mT)-O'(mT)| \le m^p \mathcal{O}(e^{-\omega/J_{\local}})$; this  immediately indicates that $D$  is indeed the effective prethermal Hamiltonian~\cite{kuwahara2016floquet,mori2016rigorous,abanin2017rigorous,abanin2017effective,machado2020long}.
%
In contrast, classical systems exhibit no such bounds---chaos causes the exponential divergence of nearby trajectories, suggesting that errors can in principle accumulate exponentially quickly.

%
%

To sharpen this intuition, we numerically explore the
Floquet dynamics of a generic classical spin model \cite{ft3}:
\begin{equation}
H_F(t)=
\begin{cases}
\sum_{i,j} J_z^{i,j} S^z_i S^z_j+\sum_{i} h_z S^z_i & 0 \le t < \frac{T}{3} \\
\sum_{i} h_y S^y_i  & \frac{T}{3} \le t < \frac{2T}{3} \\
\sum_{i,j} J^{i,j}_x S^x_iS^x_j+\sum_i h_x S^x_i &  \frac{2T}{3} \le t < T
\end{cases}
\label{Eqn:HF}
\end{equation}
where $\vec{S}_i$ is a three-dimensional unit vector.
Spin dynamics are generated by Hamilton's equations of motion $\dot{S}^\mu_i=\{S^\mu_i,H(t)\}$, using the Poisson bracket relation $\{S^\mu_i,S^\nu_j\}=\delta_{ij}\epsilon^{\mu\nu\rho}S^\rho_i$. 
The classical dynamics of an observable $O$, are then given by $O(t)=\mathcal{T} e^{\int_0^t L(t') ~dt'}[O]$, where the superoperator $L[\cdot]$ is defined by $L[\cdot]=\{\cdot, H_F \}$ \cite{fnL}. 
At lowest order in the Floquet-Magnus expansion, the static Hamiltonian is given by:
\begin{equation}
D = \frac{1}{3}\left( \sum_{i,j} J_z^{i,j} S^z_iS^z_{j} + J_x^{i,j}  S^x_iS^x_{j} + \vec{h}\cdot \vec{S}_i \right) + \mathcal{O}\left(\frac{1}{\omega}\right). 
\label{Eqn:D}
\end{equation}

To investigate the accumulation of errors, we compare the dynamics of local observables evolving under $H_F(t)$ and $D$ in a one dimensional spin chain $(N = 10^4)$ with nearest neighbor interactions~\cite{ft4}.
Deviations from the exact Floquet dynamics are measured by computing the magnetization difference between the two trajectories: $\delta M(t) = 1 - \frac{1}{N}\sum_i \vec{S}_i(t)\cdot\vec{S}'_i(t)$. 
As depicted in Fig.~\ref{fig1}(b) [top panel], $\delta M(t)$ quickly increases to a plateau value consistent with the spins in the two trajectories being completely uncorrelated; thus, $D$ cannot be thought of as the  effective prethermal Hamiltonian for $H_F(t)$. 
By contrast, the energy density remains conserved throughout the time evolution [bottom panel, Fig.~\ref{fig1}(b)], demonstrating slow Floquet heating. 

In order to pinpoint the role of chaos in the dynamics of $\delta M(t)$, we consider a slightly modified trajectory; in particular, starting with the same initial state, we first evolve under $D$ for a few Floquet cycles and then under $H_F(t)$ for all subsequent times. 
Comparing to the exact Floquet dynamics (i.e. evolution under $H_F(t)$ for all times), this protocol only differs at very early times. 
Indeed, beyond an initial, exponentially-small difference in the trajectories [arising from Eqn.~\ref{Eq:ob_c}], any additional deviation solely arises from the chaotic compounding of errors.
As depicted in Fig.~\ref{fig1}(b) [dashed curves], the magnetization difference between the modified trajectory and that of the exact Floquet dynamics, tracks $\delta M(t)$ for all times.
Crucially, this agreement demonstrates that chaos dominates the growth of $\delta M(t)$ and prevents $D$ from being the effective prethermal Hamiltonian.

\emph{Prethermal dynamics of trajectory ensembles}---While the evolution of a single trajectory cannot be captured by an effective Hamiltonian, we conjecture that $D$ captures the dynamics of \emph{ensembles} of trajectories [Fig.~\ref{fig1}(a)]; by considering an initial state composed of a region of phase space (as opposed to a single point), the details of individual chaotic trajectories become ``averaged out''.
This conjecture is made up of two separate components: (i) during the prethermal plateau, the system approaches the canonical ensemble of $D$, and (ii) $D$ accurately captures the dynamics of observables as the system evolves from local to global equilibrium.
This last component highlights the two stage approach to the prethermal canonical ensemble. First, observables on nearby sites approach the same value and the system \emph{locally} equilibrates (this occurs at time $\tau_{\mathrm{local}}$). Afterwards, the system becomes globally homogeneous as it approaches global equilibrium at time $\tau_{\mathrm{global}}$.
%

To investigate these components, we implement the following numerical experiment:
Starting from an $N=100$ spin chain, we construct an ensemble of initial states with a domain wall in the energy density at the center of the chain and study the Floquet dynamics of the local magnetization $S^z_i$ and energy density $D/N$ [Fig.~\ref{fig1}(c)]~\cite{ft5}.
%
The presence of a domain wall in the energy density enables us to distinguish between local and global equilibration. 

Focusing on the late time regime (but before Floquet heating), we find that the magnetization on opposite sides of the domain wall approaches the \emph{same prethermal plateau} [Fig.~\ref{fig1}(c)]; this precisely corresponds to the global equilibration of our spin chain. 
%
Crucially, the value of this plateau \emph{quantitatively} agrees with the mean magnetization of the corresponding canonical ensemble of $D$ calculated at the same energy density via Monte Carlo [Fig.~\ref{fig1}(c)]~\cite{SM}. 
Notably, we find agreement not only with the average value, but also with the entire distribution~\cite{SM}, thus verifying the first component of the conjecture. 

To investigate the second component, we time evolve the same ensemble of initial states for different frequencies of the drive~\cite{ft6}.
So long as $\tau_\heat \gg \tau_{\mathrm{global}}$, we find that the dynamics of local observables rapidly converge as a function of increasing frequency [Fig.~\ref{fig1}(c)].
%
%
Since the $\omega\to\infty$ limit of $H_F(t)$ precisely corresponds to Trotterized evolution under $D$, the convergence observed in Fig.~\ref{fig1}(c) indicates that $D$ is indeed the prethermal Hamiltonian for trajectory ensembles.
This is in stark contrast to the dynamics of a \emph{single} trajectory, where local observables fail to converge with increasing frequency~\cite{SM}. 


Interestingly, however, even for a single trajectory, the Floquet dynamics of either \emph{spatially} or \emph{temporally} averaged quantities are well captured by $D$.
The intuition is simple: by averaging over  different times or different spatial regions, a single trajectory effectively samples over an ensemble of different configurations [Fig.~\ref{fig1}(a)].
This insight yields a particularly useful consequence, namely, that the dynamics of a \emph{single trajectory} already encode the prethermal properties of the many-body system.

\figTwo

\emph{Prethermal dynamics with symmetry breaking}---Throughout our previous discussions, energy conservation is the only constraint that restricts the many-body dynamics within phase space.
However, symmetry-breaking can lead to additional constraints;  for example, if $D$ exhibits a discrete symmetry and this symmetry is broken at low  energy densities, then phase space is 
naturally split into multiple disjoint regions corresponding to different values of the order parameter.
As a result, the many-body dynamics under $D$ are restricted to one such region.

Floquet evolution complicates this story. %
%
In particular, one might worry that the micro-motion of the Floquet dynamics could move the system between different symmetry-broken regions of phase space.
If this were the case, prethermal symmetry-breaking phases would not be stable.
Fortunately, the ability of $D$ to approximate the dynamics over a single period (i.e.~Eqn.~\ref{Eq:ob_c}), is sufficient to constrain the Floquet evolution to a specific symmetry-broken region. 

To see this, consider, for example, a system where $D$ exhibits a discrete $\mathbb{Z}_2$ symmetry and hosts a ferromagnetic phase whose order parameter is given by the average magnetization.
%
When the energy density is below the critical value, the magnetization of the system can either be $S^z_{avg}$ or $-S^z_{avg}$. 
Given energy conservation, under a single period of evolution, the magnetization must remain the same or change sign.
%
However, Eqn.~\ref{Eq:ob_c} guarantees that the time evolved magnetization density can change, at most, by an exponentially small value in frequency.
%
This ensures that for sufficiently large driving frequencies, the magnetization cannot change sign (i.e.~move to the other symmetry-broken region) and the prethermal ferromagnet remains stable.


Crucially, symmetries of $D$ can have two different origins: they can  be  directly inherited from $H_F(t)$, or they can emerge as a consequence of the time translation symmetry of the drive~\cite{else2017prethermal,machado2020long}.
In the latter case, this can give rise to intrinsically non-equilibrium phases of matter.
To date, the study of such non-equilibrium prethermal phases has been restricted to quantum systems~\cite{lerose2019impact,mizuta2019high,lerose2019prethermal,crowley2019topological,rudner2020band,rubio2020floquet,harper2020topology,peng2021floquet,kyprianidis:2021}, where one can explicitly prove their stability~\cite{else2017prethermal,machado2020long}.
Here, we generalize and extend this analysis to classical many-body spin systems, by taking the large-$S$ limit of the quantum dynamics~\cite{mori2018floquet,SM}.

Consider a Floquet Hamiltonian which is the sum of two  terms, $H_F(t) = H_X(t) + H_0(t)$.
During a single driving period,  $H_X(t)$  generates a global rotation $X[\cdot]=\mathcal{T} e^{\int_0^T \{\cdot, H_X(t) \} dt}$, such that the system returns to itself after $M$ periods (i.e.~$X^M[\cdot] = \mathbb{I}[\cdot]$, where $\mathbb{I}$ is the identity map). 
$H_0(t)$ captures the remaining interactions in the system~
\cite{fnL}. 
For sufficiently large  frequencies, the single period dynamics (in a slightly rotated frame) are accurately captured by $X\circ e^{T\{\cdot , D \}}$, where $D$ is obtained via a Magnus expansion in the toggling frame \cite{SM}; this expansion guarantees that the dynamics generated by $D$ commute with $X$ and thus,  $X$ generates a discrete $\mathbb{Z}_M$ symmetry of the effective  Hamiltonian~\cite{else2017prethermal, machado2020long}.
Indeed, at lowest order, $D$ is simply given by the time-independent terms of $H_0(t)$ that are invariant under the global rotation.

The resulting prethermal Floquet dynamics are most transparent when analyzed at stroboscopic times $t=mT$ in the toggling frame of the $X$ rotations, wherein an observable $O$ becomes $\widetilde{O}(mT)=X^{-m}[O(mT)]$.
In this context, the dynamics of $\widetilde{O}$ are simply generated by $D$, i.e.~$\widetilde{O}(mT) = e^{mT\{\cdot , D \}}[\widetilde{O}(t=0)]$.
Thus, if the emergent $\mathbb{Z}_M$ symmetry of $D$ becomes spontaneously broken, the  system will equilibrate to a thermal ensemble of $D$ with a non-zero order parameter.


In the lab frame, the dynamics of $O$ are richer:~The global rotation changes the order parameter every period, only returning to its original value  after $M$ periods.
As a result, the system exhibits a sub-harmonic response at frequencies $1/(MT)$~\cite{else2017prethermal, machado2020long}.
This is precisely the definition of a classical prethermal discrete time crystal.

\figThree

\emph{Building a CPDTC}---Let us now turn to a numerical investigation of the classical prethermal discrete time crystal.
%
Consider the Floquet Hamiltonian in Eqn.~\ref{Eqn:HF} with an additional global $\pi$ rotation around the $\hat{x}$-axis at the end of each driving period \cite{ft7}.
At leading order, $X$ corresponds to the global $\pi$ rotation, while $D$ is given by the time averaged terms of $H_F(t)$ that remain invariant under  $X$ (i.e.~Eqn.~\ref{Eqn:D} with $h_y=h_z = 0$).
To this end, we will utilize the energy density, $D/N$, and the average magnetization, $S^z_{\ave}$,  to diagnose the prethermal dynamics and the CPDTC phase.

Let us begin by  considering a one-dimensional system with long-range interactions $J_z^{i,j}= J_z|i-j|^{-\alpha}$; when $\alpha \le 2$, $D$ exhibits ferromagnetic order below a critical temperature (or, equivalently, a critical energy density $\varepsilon_c$ which can be determined via Monte Carlo calculations)~\cite{dyson:1969,ftb}.
Taking $\alpha=1.8$ and $N=320$, we compute the Floquet dynamics starting from an ensemble with energy below $\varepsilon_c$ [Fig.~\ref{fig2}(a)]~\cite{ft8}.
After the initial equilibration to the prethermal state ($t \gtrsim \tau_{\mathrm{global}}$), the magnetization becomes homogeneous across the entire chain, signaling equilibration with respect to $D$ \cite{ft9}.
Crucially, as depicted in Fig.~\ref{fig2}(a), throughout this prethermal regime, the magnetization exhibits robust period doubling, taking on positive values at even periods and  negative values at odd periods.
%
This behavior remains stable until the CPDTC eventually ``melts'' at an exponentially late time $\tau_\mathrm{melt}$ when the energy density crosses the critical value $\varepsilon_c$ of the ferromagnetic transition of $D$ [Fig.~\ref{fig2}(a)].

Three remarks are in order.
First, because $\tau_\textrm{heat}$ is significantly longer than the interaction timescale, the system evolves between different thermal states of $D$ as it absorbs energy from the drive.
Second, the lifetime  of the CPDTC is controlled by the Floquet heating rate and thus the frequency of the drive.
Indeed, by increasing $\omega$, the lifetime of the CPDTC is exponentially enhanced, while the global equilibration time remains constant [Fig.~\ref{fig2}(b)]~\cite{SM}.
Third, we emphasize that the observed CPDTC is fundamentally distinct from period-doubling bifurcations in classical dynamical maps (e.g.~the logistic map) or the subharmonic response of a parametrically-driven non-linear oscillator~\cite{birkhoff1927dynamical,ott1981strange,strogatz2018nonlinear,kaneko1984period,kapral1985pattern,bunimovich1988spacetime,faraday1831xvii,rayleigh1883vii,rayleigh1883xxxiii,benjamin1954stability,mclachlan1947mathieu,kovacic2018mathieu,olver1982nonlinear,Arnold2009,moser1962invariant,kolmogorov1954conservation,zounes2002,yao2020classical}. 
In particular, it occurs in an \emph{isolated many-body classical system with conservative dynamics}.

Let us conclude by highlighting the central role of spontaneous symmetry breaking in observing the CPDTC. 
We do so by controlling the range of  interactions, the dimensionality, and the energy density of the initial ensemble.
To start, we consider the short-ranged version (i.e.~nearest neighbor interactions) of the 1D classical spin chain discussed above.  
Without long-range interactions, 
ferromagnetic order is unstable at any finite temperature~\cite{Landau1937}, and this immediately precludes the existence of a CPDTC. 
This is indeed borne out by the numerics [Fig.~\ref{fig2}(c)]: We observe a fast, frequency-independent decay of the magnetization to its infinite-temperature value.

While nearest-neighbor interactions cannot stabilize ferromagnetism in 1D, they do so in higher dimension. 
To this end, we explore the same Floquet model (i.e.~Eqn.~\ref{Eqn:HF}) on a two dimensional square lattice.
For sufficiently low energy densities, the system equilibrates to a CPDTC phase [Fig.~\ref{fig3}(a)], while above the critical temperature, the system equilibrates to a trivial phase [Fig.~\ref{fig3}(b)].
We hasten to emphasize that our framework is not restricted to the period-doubled ($M=2$) CPDTC and it immediately ports over to more general notions of time crystalline order, including both  higher-order  ($M>2$) and \emph{fractional} CPDTCs (see supplemental material for additional numerics)~\cite{SM,pizzi2021higher}.

Our work opens the door to a number of intriguing directions. 
First, it would be interesting to explore the generalization of classical prethermal time crystals to quasi-periodic driving \cite{else2020long}.
Second, although we have presented extensive numerical and analytic evidence for the presence of an effective Hamiltonian (for trajectory ensembles), sharpening our analysis into a proof would provide additional insights in the nature of many-body classical Floquet systems.

\begin{acknowledgments}
We gratefully acknowledge discussions with and the insights of Ehud Altman, Marin Bukov, Soonwon Choi, Wade Hodson, Chris Jarzynski, Canxun Zhang, and Chong Zu.
This work is supported in part by the NSF (Grant No.~PHY-1654740 and the QLCI program through Grant No.~OMA-2016245), the DARPA DRINQS program  (Grant No.~D18AC00033), the A.~P.~Sloan foundation, the David and Lucile Packard foundation,  and the W. M. Keck Foundation.
\end{acknowledgments}

\emph{Note added:} During the completion of this work, we became aware of complementary work exploring prethermal phases of matter in classical spin systems \cite{pizzi2021classical}.

\bibliography{references}

\begin{thebibliography}{88}
\bibitem [{ft1()}]{ft1}%
  \BibitemOpen
  \href@noop {} {\bibinfo  {journal} {More precisely, we note that the
  classical Floquet dynamics are generated by the superoperator $L(t)[ \cdot] =
  \{ \cdot, H_F(t)\}$. The time evolution operator over a single period is then
  given by $U_F = \mathcal{T} e^{\int_0^T L(t) dt} \equiv e^{L_F T}$, where
  $L_F$ is a time-independent superoperator. The static Hamiltonian,
  $\mathcal{H}_F$, corresponding to $L_F$ is then given by $L_F = \{\cdot,
  \mathcal{H}_F\}$. The Floquet-Magnus expansion constructs $\mathcal{H}_F$
  order-by-order in $\frac{J_{\local}}{\omega}$, i.e.~$\mathcal{H}_F =
  \sum_{n=0}^{\infty} \left (\frac{J_{\local}}{\omega} \right )^n D^{(n)}$,
  where $D^{(n)}$ is the $n^{\textrm{th}}$ order term of the expansion. Note
  that the effective static Hamiltonian, $D$, is then defined as $D =
  \sum_{n=0}^{n^*} \left (\frac{J_{\local}}{\omega} \right )^n D^{(n)}$}\
  }\BibitemShut {NoStop}%
\bibitem [{ft2()}]{ft2}%
  \BibitemOpen
\bibfield  {journal} {  }\href@noop {} {\bibinfo  {journal} {Assuming that the
  system is extensive and power-law light-cones exist as defined via
  Lieb-Robinson bounds}\ }\BibitemShut {NoStop}%
\bibitem [{ft3()}]{ft3}%
  \BibitemOpen
\bibfield  {journal} {  }\href@noop {} {\bibinfo  {journal} {Throughout this
  work, we utilize the following generic set of parameters
  $\{J_z,J_x,h_x,h_y,h_z\}=\{-1.0,0.79,0.17,0.23,0.13\}$}\ }\BibitemShut
  {NoStop}%
\bibitem [{fnL()}]{fnL}%
  \BibitemOpen
\bibfield  {journal} {  }\href@noop {} {\bibinfo  {journal} {We note that the
  multiplication of the superoperators (functions of observables) should be
  understood as function composition. In particular, $(L_1\circ L_2)[\cdot ] =
  L_1[L_2[\cdot ]]$. The $n^\mathrm{th}$ power of $L$ is then defined
  inductively by $L^n=L\circ L^{n-1}$. Therefore, evolving under the Floquet
  Hamiltonian for $m$ periods, an observable becomes $O(mT)=\left( \mathcal{T}
  e^{\int_0^T L(t) dt}\right)^m[O]$. In a similar fashion, let us define
  $X^{-1}$ as the inverse of the map $X$}\ }\BibitemShut {NoStop}%
\bibitem [{ft4()}]{ft4}%
  \BibitemOpen
\bibfield  {journal} {  }\href@noop {} {\bibinfo  {journal} {In general, the
  chaotic nature of $D$ means that numerically integrating the equations of
  motion to later times requires exponentially better precision, making the
  numerical treatment very difficult. By contrast, each term of $H_F(t)$
  corresponds to a precession of the spins along one of three axis, which can
  be straightforwardly analytically calculated without resorting to numerical
  integration methods}\ }\BibitemShut {NoStop}%
\bibitem [{ft5()}]{ft5}%
  \BibitemOpen
\bibfield  {journal} {  }\href@noop {} {\bibinfo  {journal} {We initialize each
  spin along either $+\hat{z}$ or $-\hat{z}$ direction. By tuning the number of
  domain walls, we can control the local energy density of the system. While
  the spins on the right half of the chain are initialized in a completely
  ferromagnetic state, the spins on the left half repeat the following pattern:
  $\downarrow \uparrow \downarrow$. Therefore, the energy density across the
  chain exhibits a domain wall at the center of the chain. To bring out the
  ensemble effect, we add small random noise to the azimuthal angle and perform
  an average of the subsequent dynamics over these slightly different initial
  states}\ }\BibitemShut {NoStop}%
\bibitem [{ft6()}]{ft6}%
  \BibitemOpen
\bibfield  {journal} {  }\href@noop {} {\bibinfo  {journal} {The local
  equilibration time $\tau_{\local}$ corresponds to the time when nearby spins
  approach the same value. To identify $\tau_{\local}$, we measure the time
  when $S^z_{43}$ and $S^z_{45}$, initially pointing in opposite directions,
  exhibit the same magnetization}\ }\BibitemShut {NoStop}%
\bibitem [{ft7()}]{ft7}%
  \BibitemOpen
  \href@noop {} {\bibinfo  {journal} {The $\pi$ rotation can be generated by a
  magnetic field along the $\hat{x}$ direction for time $\tau_\pi$ with
  strength $\pi/2\tau_\pi$. Since the dynamics are independent of the value of
  $\tau_\pi$, we consider it to be an instantaneous rotation, $\tau_\pi \to 0$.
  \\ In order to better highlight the role of $D$ throughout the prethermal
  regime, we consider instead a symmetrized version of $H_F(t)$ such that its
  first-order contribution in inverse frequency is zero and thus the dynamics
  are better captured by the zero-th order terms. We note that this does not
  change the results qualitatively}\ }\BibitemShut {NoStop}%
\bibitem [{ftb()}]{ftb}%
  \BibitemOpen
  \href@noop {} {\bibinfo  {journal} {By performing extensive Monte Carlo
  simulations, we obtain the critical energy density $\epsilon_c\approx-0.53$
  for the transition \cite{SM}}\ }\BibitemShut {NoStop}%
\bibitem [{ft8()}]{ft8}%
  \BibitemOpen
\bibfield  {journal} {  }\href@noop {} {\bibinfo  {journal} {To prepare the
  initial state ensemble, we first start with the fully polarized system, flip
  every tenth spin, and then add a small amount of noise to the azimuthal
  angle. The energy density of the resulting ensemble is $\varepsilon =
  -1.10$}\ }\BibitemShut {NoStop}%
\bibitem [{ft9()}]{ft9}%
  \BibitemOpen
\bibfield  {journal} {  }\href@noop {} {\bibinfo  {journal} {We note that the
  fluctuations of $S^z_i$ are local thermal fluctuations, and they decrease as
  we average over more realizations}\ }\BibitemShut {NoStop}%
\end{thebibliography}

\end{document}


\title{Supplementary Material:\\
Floquet phases of matter via classical prethermalization}
\author{Bingtian Ye}
\affiliation{Department of Physics, University of California, Berkeley, CA 94720, USA}
\author{Francisco Machado}
\affiliation{Department of Physics, University of California, Berkeley, CA 94720, USA}
\affiliation{Materials Science Division, Lawrence Berkeley National Laboratory, Berkeley, CA 94720, USA}
\author{Norman Y. Yao}
\affiliation{Department of Physics, University of California, Berkeley, CA 94720, USA}
\affiliation{Materials Science Division, Lawrence Berkeley National Laboratory, Berkeley, CA 94720, USA}

\maketitle
\section{Proof for prethermalization with the presence of symmetries}
In this section we prove that the prethermal effective Hamiltonian can exhibit an emergent symmetry (protected by the time translation symmetry), while preserving the remaining properties of the prethermal regime, namely, exponentially slow heating (Eqn.~1 of the main text) and an approximate description of the dynamics at the single trajectory level for a single period of the drive (Eqn.~2 of the main text).
This generalization follows closely the quantum results presented in Refs.~\cite{else2017prethermal, machado2020long} using the machinery first introduced in Ref.~\cite{mori2018floquet} to extend it to classical systems.
More specifically, we treat the classical spin as the large-$S$ limit of the quantum spin model; for arbirtrarily large $S$, each spin-$S$ degree of freedom can be separated into multiple spin-1/2 degrees of freedom. 
This approach allows one to immediately translate the quantum results to the corresponding classical system. 

Let us begin by reviewing the quantum case.
Without loss of generality, we consider a quantum spin-$S$ system with one-body and two-body terms:
\begin{equation}
    \hat{H}_F(t) =  \underbrace{g(t)\sum_i^N \hat{S}^x_i}_{\hat{H}_X(t)}+\underbrace{\frac{1}{2S}\sum_{ij}^N\sum_{\mu \nu=x,y,z} J^{ij}_{\mu\nu}(t)\hat{S}^\mu_i \hat{S}^\nu_j+\sum_i^N\vec{h}^i(t)\cdot \hat{\vec{S}}_i}_{\hat{H}_0(t)},
\label{eq:general_H}
\end{equation}
where $g(t)$, $J^{ij}(t)$ and $\vec{h}^i(t)$ are periodic functions of time with a period of $T$, and the dynamics are generated by the Heisenberg equations of motion for $\hat{S}^\mu_i$, or equivalently the Schr\"{o}dinger equation for the quantum state. 
Here, we assume that the interactions are not extensive, i.e. $\sum_{j}\sum_{\mu\nu} \max\limits_{t} |J^{ij}_{\mu\nu}(t)|$ is finite for all $i$. 
Crucially, we assume that $g(t)$ satisfies $M\int_0^T g(t)dt=2k\pi$, where $M$ and $k$ are two coprime integers. 
This means that, under the dynamics generated by $\hat{H}_X(t)$, the system will return to the same state every $M$ driving cycles,  i.e.~$\hat{X}^M=\left[ \mathcal{T} e^{-i\int_0^T \hat{H}_X(t)dt}\right]^M= \mathds{1}$. 
The corresponding maximal local energy scale is then defined as
\begin{equation}
J_{\mathrm{local}}:=\max_{\mathrm{site}~i} \sum_{A:i\in A} \left \|\hat{c}^{A} \right \| ,
\label{eq:Jlocal}
\end{equation}
where $\hat{c}^A$ correspond to local terms in $\hat{H}_0 = \sum_A \hat{c}^A$ with support on the set of spins $A$, and $\|\cdot\|$ is the operator norm. 

In this particular model of Eqn.~\ref{eq:general_H}, the local terms are:
\begin{equation}
\begin{cases}
    \hat{c}^A = \frac{1}{S}\sum_{\mu \nu=x,y,z} J^{ij}_{\mu\nu}(t)\hat{S}^\mu_i \hat{S}^\nu_j, \qquad  \mathrm{for~} A=\{i,j\};\\
    \hat{c}^A = \vec{h}^i(t)\cdot \hat{\vec{S}}_i, \qquad\qquad\qquad\qquad\qquad \mathrm{for~} A=\{i\}.\\
\end{cases}
\label{eq:ca}
\end{equation}

\begin{figure}[t]
    \includegraphics[width=3.0in]{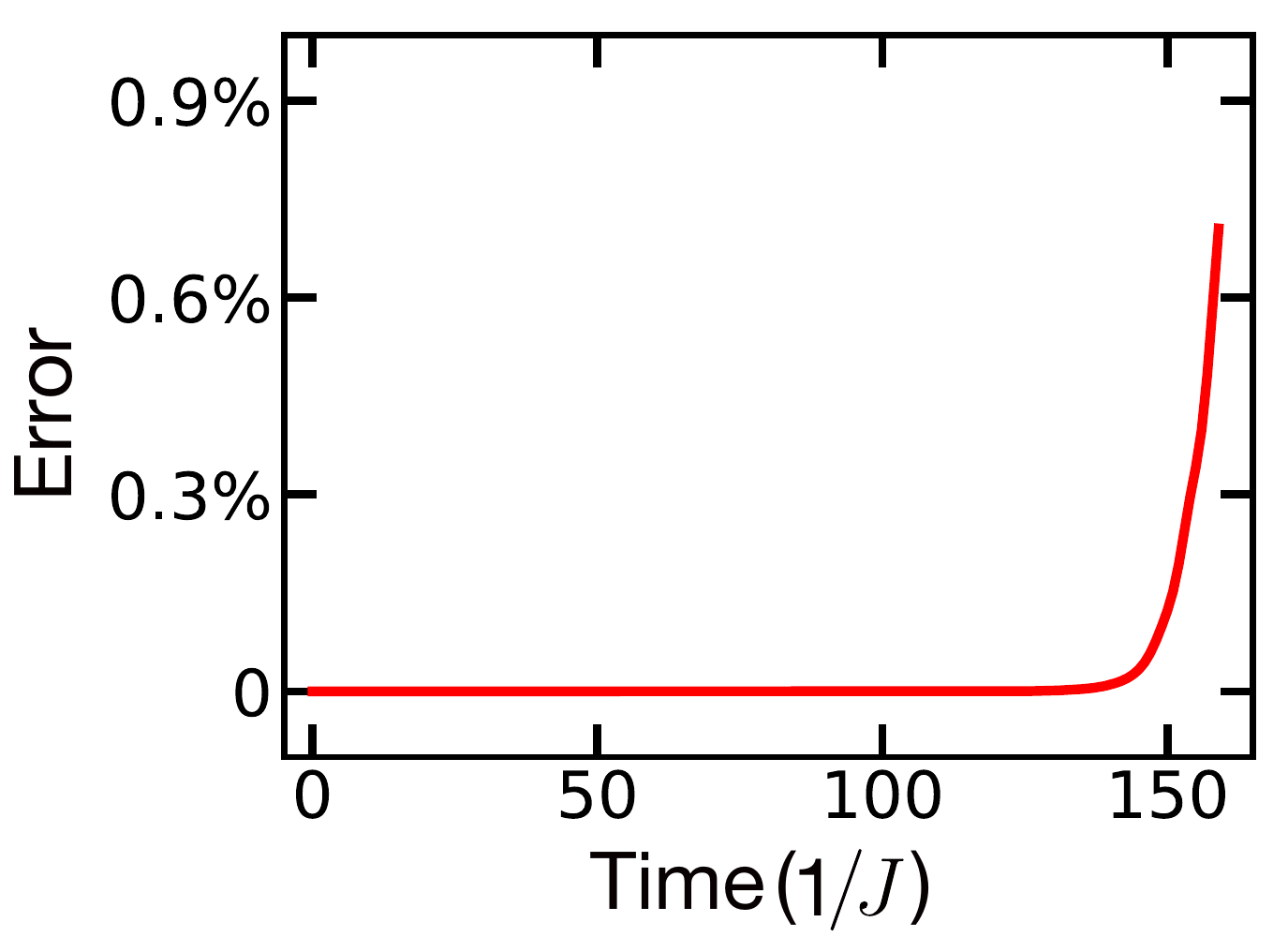}
    \centering
    \caption{
    Numerical errors as a function time. To estimate the error, we evolve the system with two different step lengths ($1/800J$ and $1/1600J$) in the fourth-order Runge-Kutta method, and compute the difference between the two resulting dynamics. The error is defined as their magnetization difference.
    }
    \label{fig:Runge-Kutta}
  \end{figure}

Crucially, it is proven that, when the frequency of the drive is much larger than the local energy scale, i.e:
\begin{equation}
J_{\mathrm{local}} \ll \frac{1}{MT}, \label{eq:condition}
\end{equation}
there exist a prethermal Hamiltonian $D$ and a (slightly rotated) symmetry operation $\hat{X}_{\mathrm{rot}}\approx \hat{X}$ such that \cite{else2017prethermal,machado_exponentially_2017}:
\begin{align}
    \hat{X}_{\mathrm{rot}}^M&=\mathds{1}, \label{eq:quantum_bound}\\ 
    [\hat{D},\hat{X}_{\mathrm{rot}}]&=0,\\
    \frac{1}{N}\|\hat{D}(mT)-\hat{D}(0)\|&< c_1\cdot mJ_{\mathrm{local}}\cdot\mathcal{O}(e^{-\omega/J_{\mathrm{local}}}),\\
    \|\hat{O}(T)-\hat{X}_{\mathrm{rot}}^{-1}\hat{O}'(T)\hat{X}_{\mathrm{rot}}\|&< c_2 \cdot \|\hat{O}\|\cdot\mathcal{O}(e^{-\omega/J_{\mathrm{local}}}), \label{eq:local_Error}
\end{align}
where $\hat{O}$ is any local observable and $c_1,c_2$ are constants independent of frequency and $J_{\mathrm{local}}$.
We note that $\hat{O}'(T)$ corresponds to operator $\hat{O}$ evolved under $\hat{D}$ for time $T$, i.e. $\hat{O}'(T)=e^{i\hat{D}T}\hat{O}(0)e^{-i\hat{D}T}$. 
When the system has a polynomial Lieb-Robinson bound, the error of the operator (Eqn.~\ref{eq:local_Error}) grows algebraically with time and the dynamics remain well approximated by $\hat{D}$ for an exponentially long time in the frequency of the drive.

Moving on to the classical case, the corresponding classical spin system can be treated using Eqn.~\ref{eq:general_H} by taking $S\rightarrow \infty$.
Correspondingly, the classical spin variable (i.e. a unit vector) can be viewed as $\vec{S}=\hat{\vec{S}}/S$ \cite{mori2018floquet}. 
However, the results for finite $S$ in Eqn.~\ref{eq:quantum_bound} cannot be immediately applied to the classical case, since the local energy scale diverges, $J_{\mathrm{local}}=||\hat{H}_0||\propto S$, which invalidates the high-frequency condition (Eqn.~\ref{eq:condition}). 
Fortunately, previous work introduced a mathematical treatment of the Hamiltonian that solved this precise problem~\cite{mori2018floquet}. 
The main idea is as follows: We start by decomposing each spin $\hat{\vec{S}}_i$ into $2S$ spin-1/2 operators $\{\hat{\vec{s}}_{i,a}\}$:
\begin{equation}
    \hat{S}^\mu_i=\sum^{2S}_{a=1} \hat{s}^\mu_{i,a}.
\end{equation}
With this substitution, the two parts of the Floquet Hamiltonian become:
\begin{equation}
\begin{split}
    \hat{H}_X(t) &= g(t)\sum_{i,a} \hat{s}^x_{i,a}\\
    \hat{H}_0(t) &= \frac{1}{2S}\sum_{i,a,j,b}^N\sum_{\mu \nu=x,y,z} J^{ij}_{\mu\nu}\hat{s}^\mu_{i,a} \hat{s}^\nu_{j,b}+\sum_i\vec{h}_i(t)\cdot \hat{\vec{s}}_{i,a}. 
\end{split}
\label{eq:split_H}
\end{equation}
From the perspective of the spin-1/2, $\hat{H}_X$ still respects its distinguishing property: after $M$ cycles of evolution under $\hat{H}_X$ alone, each individual spin-1/2 returns to the initial state. 
Meanwhile, the local energy scale of $\hat{H}_0$ remains finite:
\begin{equation}
     \frac{1}{S}\sum_{j}\sum_{b=1}^{2S}\sum_{\mu \nu} \vert J^{ij}_{\mu\nu}\vert \cdot\|\hat{s}^\mu_{i,a} \|\cdot\|\hat{s}^\nu_{j,b}\|+\sum_{\mu}|h^{\mu}_i(t)|\cdot \|\hat{s}^{\mu}_{i,a}\|     \le \frac{1}{2}\sum_{j}\sum_{\mu \nu} \vert J^{ij}_{\mu\nu}\vert +\frac{1}{2}\sum_{\mu}|h^{\mu}_i(t)|.
\label{eq:split_H}
\end{equation}
Therefore, as this construction holds for any $S$, one can then safely take the $S\to\infty$ limit and extend the conclusions of quantum derivation (Eqns.~\ref{eq:quantum_bound}-\ref{eq:local_Error}) to classical systems, immediately proving Eqn.~1 and Eqn.~2 in the main text for the prethermal dynamics of \emph{classical systems} with an \emph{emergent symmetry}.

Here, we remark that by splitting the large-$S$ spins into $2S$ spin-1/2 degrees of freedom, one reduces the local energy scale of each spin degrees at the expense of generating an additional dimension (where position is labeled by the index of the spin-1/2) and all-to-all coupling along this virtual dimension. 
Although these very long-range interactions do not lead to the divergence of the local energy scale, it precludes the notion of locality and thus meaningful Lieb-Robinson---as a result, $\hat{D}$ cannot reproduce the dynamics over multiple Floquet cycles.

Let us end this section by formalizing these results in the language of classical dynamics. 
In classical spin systems, the dynamics are generated by Hamilton's equations of motion $\dot{S}^\mu_i=\{S^\mu_i,H(t)\}$, using the Poisson bracket relation $\{S^\mu_i,S^\nu_j\}=\delta_{ij}\epsilon^{\mu\nu\rho}S^\rho_i$. 
Equivalently, at any time $t$, any observable $O(t)$ can thought of a function of the initial values of all observables.
This can be expressed by formally integrating the equations of motion: $O(t)=\mathcal{T} e^{\int_0^t L(t') dt'}[O]$, where the superoperator $L[\cdot]$ is defined by $L[\cdot]=\{\cdot, H_F \}$. 
We note that the multiplication of the superoperators (functions) should be understood as function composition, i.e.~$(L_1\circ L_2)[\cdot ] = L_1[L_2[\cdot ]]$. 
Correspondingly, the $n^{\mathrm{th}}$ power of $L$ is then defined inductively by $L^n=L\circ L^{n-1}$, and the exponential function of the superoperators is naturally defined by its Taylor expansion.

With such notations for classical dynamics, the properties of the classical prethermalization can be stated as the follows. 
Consider a classical spin system with a Hamiltonian consisting of two parts:
\begin{equation}
    H_F(t) = H_X(t)+H_0(t),
\end{equation}
where $H_X(t)$ generates a global rotation $X=\mathcal{T} e^{\int_0^T \{\cdot, H_X(t)\}dt}$ over a single period such that the system returns to itself every $M$ periods, i.e.~$X^M[\cdot] = \mathbb{I}[\cdot]$. 
If the local energy scale of $H_0(t)$ is sufficiently small:
\begin{equation}
J_{\mathrm{local}}:=\max_{\mathrm{site}~i}\sum_{A:i\in A}\left \|c^{A}\right\|\ll \frac{1}{MT},
\end{equation}
where $c_A$ corresponds to classical analogue of $\hat{c}_A$ in Eqn.\ref{eq:ca} (i.e. local terms with support in $A$), 
then there exists a prethermal Hamiltonian $D$ and a (slightly) tilted symmetry operation $X_{\mathrm{rot}}[\cdot]\approx X[\cdot]$ such that
\begin{align}
    X_{\mathrm{rot}}^M[\cdot] &= \mathbb{I}[\cdot],\\
    X_{\mathrm{rot}}[D]&=D,\\
    \frac{1}{N}|D(mT)-D(0)|&< c_1.\cdot mJ_{\mathrm{local}}\cdot\mathcal{O}(e^{-\omega/J_{\mathrm{local}}}),\\
    \left \|O(T)-X_{\mathrm{rot}}[O'(T)] \right\|&< c_2.\cdot |O|\cdot\mathcal{O}(e^{-\omega/J_{\mathrm{local}}}),
\end{align}
where $c_1,c_2$ are constants independent of frequency and $J_{\mathrm{local}}$, $O$ is any classical local observable evolving under $H_F(t)$, and $O'(T)$ corresponds to the observable $O$ evolved under $D$ for time $T$, i.e.~$O'(T)=e^{T\{\cdot,D\}}[O(0)]$.

\begin{figure}[t]
    \includegraphics[width=3.0in]{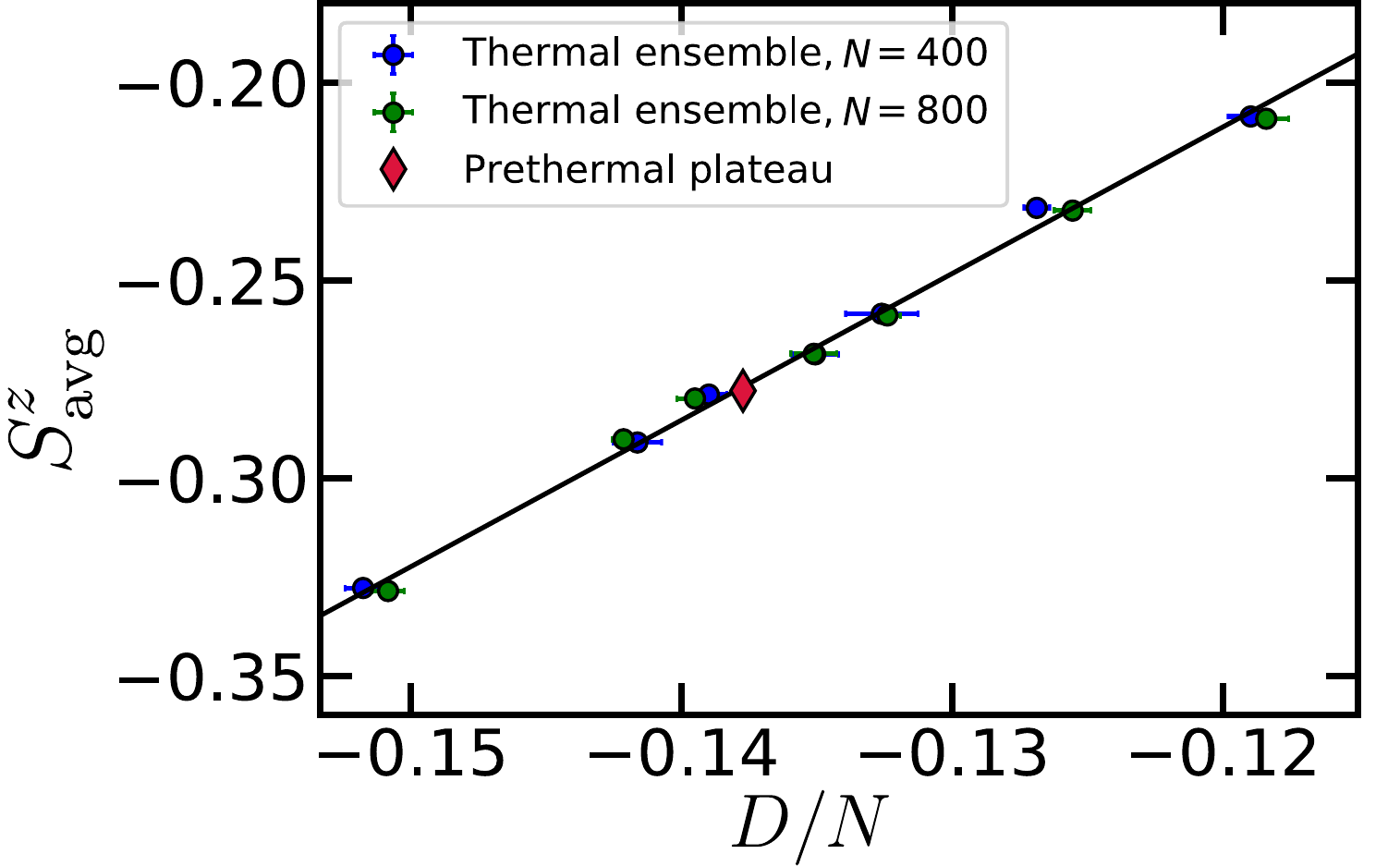}
    \centering
    \caption{The average magnetization $S^z_{\mathrm{avg}}$ as a function of the energy density $D/N$. To calculate $S^z$ and $D/N$ of a thermal ensemble with respect to $D$, we perform classical Monte Carlo simulations at different temperatures. The prethermal plateau value is obtained by directly compute the observables of the fast-driving Floquet system in the prethermal regime. Crucially, the prethermal plateau precisely agrees with the thermal ensemble of $D$. 
    }
    \label{fig:MonteCarlo_plateau}
  \end{figure}

\section{Numerical comparison between Floquet dynamics and the effective description}
In the main text, we numerically compare the dynamics under the true Floquet evolution and the effective Hamiltonian. 
At short times, we consider the evolution of a single state under the two Hamiltonians, while at long times, we compare the Floquet prethermal state to the corresponding thermal state of the effective Hamiltonian. 
Since a discussion around the numerical simulation of the true Floquet dynamics $H_{F}$ is already presented in the main text, in this section we provide additional details on the numerics associated with the effective Hamiltonian $D$. 

\subsection{Benchmark the dynamics under the effective Hamiltonian}
For short times $\sim 10^2/J$, we simulate the dynamics under $D$ by numerically solving the equations of motion $\dot{S}^\mu_i=\{S^\mu_i,D\}$. 
Since $D$ involves the terms along all three axes, one cannot utilize the techniques for the Floquet evolution that enable the very long-time simulation \cite{howell2019asymptotic}. 
Instead, we solve the non-linear problem by using a fourth-order Runge–Kutta method, where the precision of the simulation is controlled by the step length. 
As shown in Fig.~\ref{fig:Runge-Kutta}, we confirm that, within the timescale considered ($\sim 150/J$), a time step of $\sim 1/1000J$ guarantees that the numerical error is less than $1\%$.

\subsection{Monte Carlo results for the equilibrium states}
In the main text, we present two important conclusions regarding the equilibration of the prethermal dynamics.
First, we show that, during the prethermal regime, the system approaches the thermal state of the effective Hamiltonian. 
Second, we demonstrate that the time crystalline order decays when the temperature of the system crosses the critical temperature of the phase transition of the underlying effective Hamiltonian $D$. 
In order to reach these two conclusions, we compute the equilibrium states with respect to the static effective Hamiltonian, which we achieve by performing a classical Monte Carlo simulation. 
In this section, we provide additional details and data regarding this calculation.

During a Monte Carlo simulation, the system is allowed to transition between different classical configurations according to rates determined by energy difference between the configuration and the temperature. 
By choosing these rates to obey detailed balance and ergodicity properties (we make use of the Metropolis algorithm), the ensemble of configurations explored matches the equilibrium ensemble and thermal averages can be computed.

For each run of our Monte Carlo simulation, we start from a completely random initial configuration and then run the algorithm for $10000$ steps to approach equilibrium; we then build the necessary statistics using the ensemble generated from the following $30000$ steps of the algorithm.
For each data point, we perform the simulation with $8$ different random initial states, and consider the average across the $8$ runs and their uncertainty. 

To compare the Floquet prethermal plateau with the thermal ensemble of $D$, we compute the average magnetization $S^z_{\mathrm{avg}}$ at a function of the energy density $D/N$ (Fig.~\ref{fig:MonteCarlo_plateau}). 
Crucially, we also plot the observed prethermal plateau value in Fig.~\ref{fig:MonteCarlo_plateau}, and observe great agreement with the thermal ensemble value. 

\begin{figure}[t]
    \includegraphics[width=5.5in]{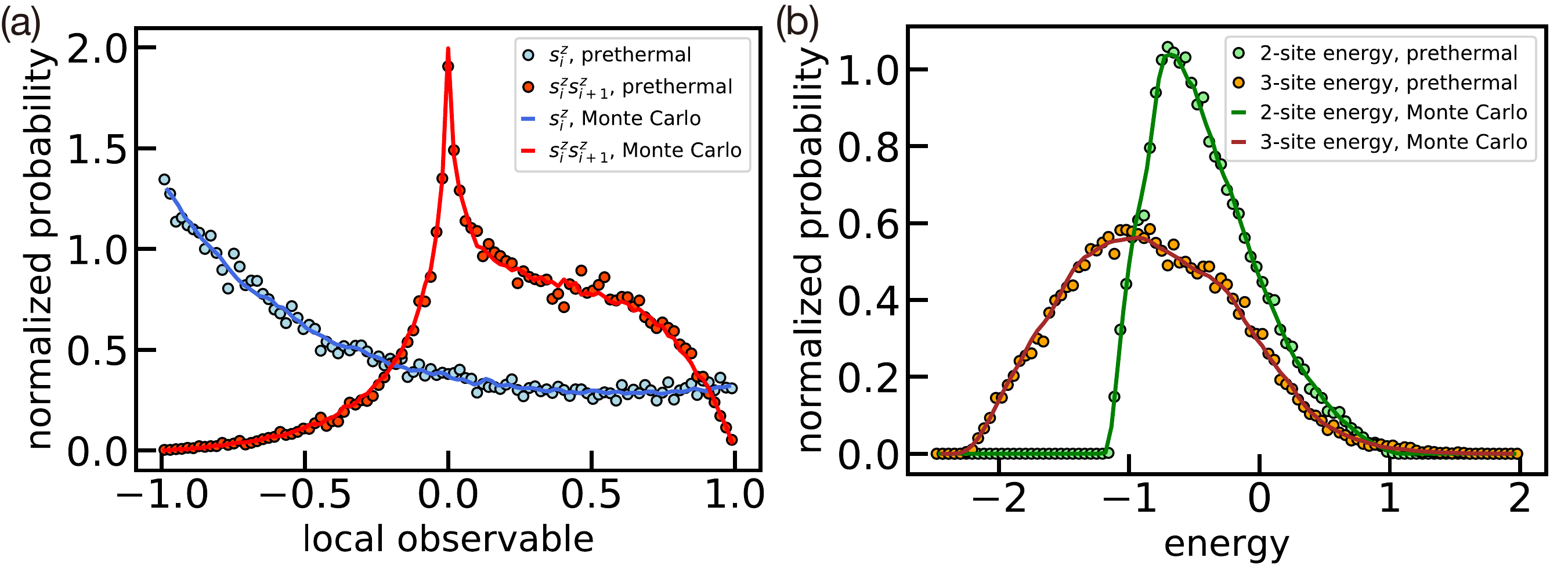}
    \centering
    \caption{Probability distribution of (a) local observables and (b) energies corresponding to a Floquet prethermal state and a canonical ensemble. The 2-site and the 3-site energy refers to the energy measured on 2 and 3 consecutive sites, respectively. The canonical ensemble is obtained via Monte Carlo simulation. For both the local observables and the energy, the distributions of the Floquet prethermal state correponding to $H_F(t)$ perfectly agree with that of the true thermal state corresponding to $D$.
    }
    \label{fig:distribution}
  \end{figure}

Beyond the mean value, the complete description of an ensemble also includes its fluctuations. 
Therefore, to fully verify that the system approaches the thermal ensemble of $D$ in the prethermal regime, we further study the full distribution of the local observables. 
In Fig.~\ref{fig:distribution}(a), for the distributions of both the single-site and the two-site observables (i.e. local magnetization and its correlation), we observe excellent agreement between the Floquet prethermal state and the thermal (canonical) ensemble of $D$. 
In Fig.~\ref{fig:distribution}(b), the local energy also exhibit the same great agreement. 
Such agreement highlights that the Floquet prethermal state is indeed well approximated by the thermal canonical state of $D$.

\begin{figure}[t]
    \includegraphics[width=3.5in]{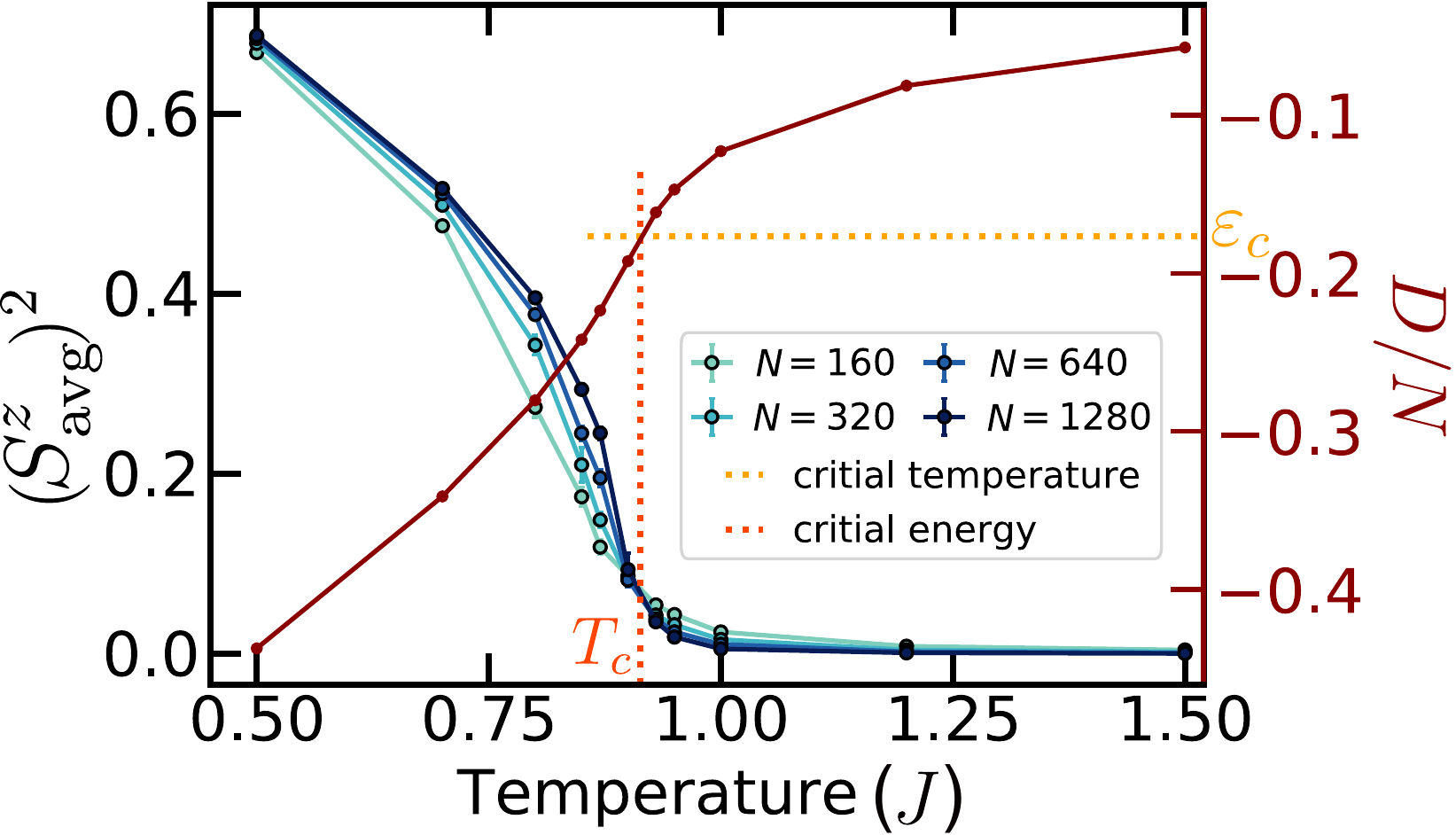}
    \centering
    \caption{
    The ferromagnetic order parameter $(S^z_{\mathrm{avg}})^2$ and the energy density $D/N$ as a function of the temperature, calculated by Monte Carlo simulation. 
    As the system size $N$ increases, the transition of the order parameter from zero to non-zero becomes sharper. 
    We identify the critical temperature as the crossing point of the curves with different system sizes. 
    The energy density as a function of temperature then allows us to obtain the critical energy density $\varepsilon_c$, which we use to compare against the melting point of the CPDTC. 
    }
    \label{fig:MonteCarlo_critical}
  \end{figure}

To confirm that CPDTC melts when its temperature [energy density] matches the critical point of the symmetry breaking phase transition of $D$, we studied the ferromagnetic order parameter $(S^z_{\mathrm{avg}})^2$ as a function of temperature. 
By performing the finite size scaling, we obtain the critical temperature and, thus, the corresponding critical energy (Fig.~\ref{fig:MonteCarlo_critical}). 

\section{Dynamics of local observables in different models}
In the main text, we studied the dynamics of local magnetization in the low-temperature long-range interacting model. 
Such dynamics are crucial to understand when the system has reached prethermal equilibrium, which is necessary for claiming the observation of a CPDTC, as the prethermal phase is well-defined after any transient dynamics associated with equilibration decay. 
Moreover, we demonstrated that the single site magnetization agrees with the global magnetization afterwards, indicating that the subharmonic oscillation can be diagnosed by local quantities without performing an spatial average across the entire system. 
In this section, we show additional data on the local dynamics of the other models studied in the main text: the short-range interacting 1D spin chain, and the short-range interacting 2D spin model at both low and high temperatures. 

\begin{figure}[t]
    \includegraphics[width=6in]{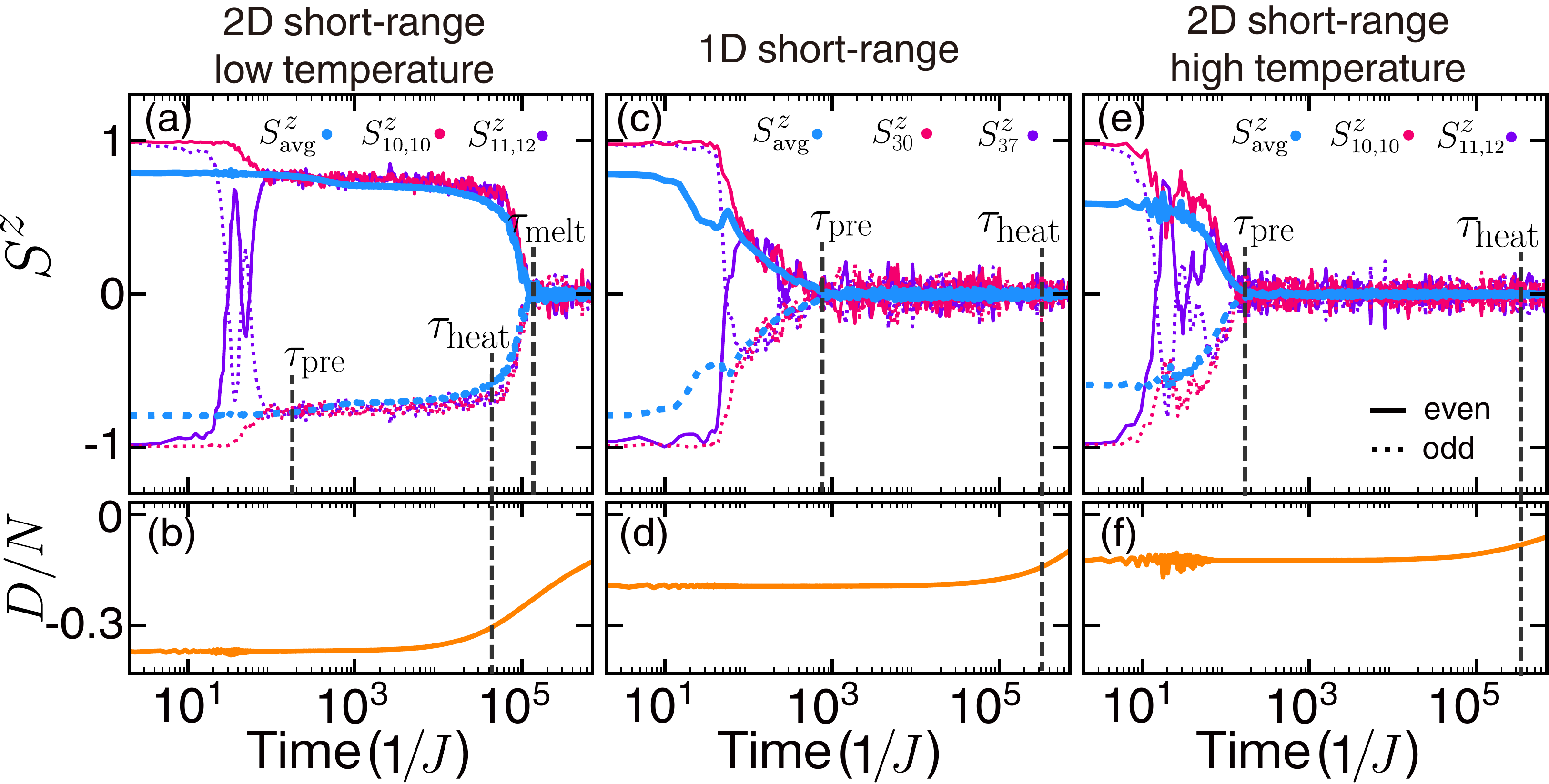}
    \centering
    \caption{
    Magnetization dynamics [(a)(c)(e)] and energy [(b)(d)(f)] evolution of various Floquet spin systems: (a)(b) short-range interacting spins on a 2D square lattice at low temperature; (c)(d) short-range interacting spins on a 1D chain; (e)(f) short-range interacting spins on a 2D square lattice at high temperature. 
    In all cases, the local magnetization approaches the global average after equilibration with respect to the effective Hamiltonian. However, only the low-temperature 2D system exhibits subharmonic oscillation in both the global and local magnetization until exponentially late heating timescales. 
    }
    \label{fig:CPDTC_Diff}
  \end{figure}

Among all these cases, only the short-range interacting 2D model at low temperature hosts time-crystalline order.
Indeed, its local dynamics exhibit qualitatively the same behaviors as the low-temperature long-range interacting 1D spin chain. 
In Fig.~\ref{fig:CPDTC_Diff}, starting from an initial ensemble with low energy density, the local magnetization on different sites first follows different time traces, but eventually approach the same value, which equals the globally averaged magnetization and agrees with the prethermal equilibrium. 
Then the local magnetization exhibits a robust oscillation between positive and negative values.
By contrast, the lack of a symmetry broken phase in the other models prevents a stable time-crystalline phase.
Accordingly, we observe similar pattern in the local equilibration process, but with the local magnetization approaches zero (Fig.~\ref{fig:CPDTC_Diff}). 
This shows that there are no non-trivial prethermal phase: not only the globally averaged quantities but also the local quantities trivially remains constant throughout the prethermal equilibrium regime. 

\section{Different Classes of Prethermal Time Crystals}
In the main text, we show a CPDTC phase with a period of $2T$ in both one and two dimensions. 
Actually, our framework immediately suggests that by tuning the symmetry operation $X$, one can obtain $D$ with different emergent symmetries, and further realize CPDTC with different periods. 
Here in this section, we provide two such examples: 
1) a 3-DTC with a period of $3T$ and 
2) a fractional DTC with a period of $\frac{5}{2}T$. 

In particular, we focus on the 1D long-range interacting spin chains, which allows a spontaneously broken symmetry phase at low temperatures. 
However, if the effective Hamiltonian only includes two-body interactions, then the existence of any $\mathbb{Z}_M$ symmetry with $M>2$ will actually imply that the system has a $U(1)$ symmetry. 
Therefore, in order to introduce a $\mathbb{Z}_M$ symmetry while avoiding the continuous $U(1)$ symmetry, we add a nearest-neighbor three-body term ($J_{zz} S^z_{i-1} S^z_i S^z_{i+1}$) in the Hamiltonian, and keep other two-body terms the same as Eqn.~4 in the main text. 
Another important modification to the Floquet Hamiltonian is that instead of having a $\pi$-rotation around the $\hat{x}$-axis, we evolve the system with a $\frac{2\pi}{M}$-rotation at the end of each driving period.
More specifically, the Floquet Hamiltonian is written as:
\begin{equation}
H_F(t)=
\begin{cases}
\sum_{i,j} \frac{J_z}{|i-j|^\alpha} S^z_i S^z_j+\sum_{i} J_{zz} S^z_{i-1} S^z_i S^z_{i+1}+\sum_{i} h_z S^z_i & 0 \le t < \frac{T}{6}, \frac{5T}{6} \le t < T \\
\sum_{i} J_y S^y_iS^y_{i+1}+\sum_i h_y S^y_i & \frac{T}{6} \le t < \frac{T}{3}, \frac{2T}{3} \le t < \frac{5T}{6} \\
\sum_{i} J_x S^x_iS^x_{i+1}+\sum_i h_x S^x_i &  \frac{T}{3} \le t < \frac{2T}{3}\\
g\sum_i S^x_i &  T \le t < T+T'
\end{cases}
\end{equation}
where $gT'=\frac{2\pi}{M}$. 

\begin{figure}[t]
    \includegraphics[width=5.5in]{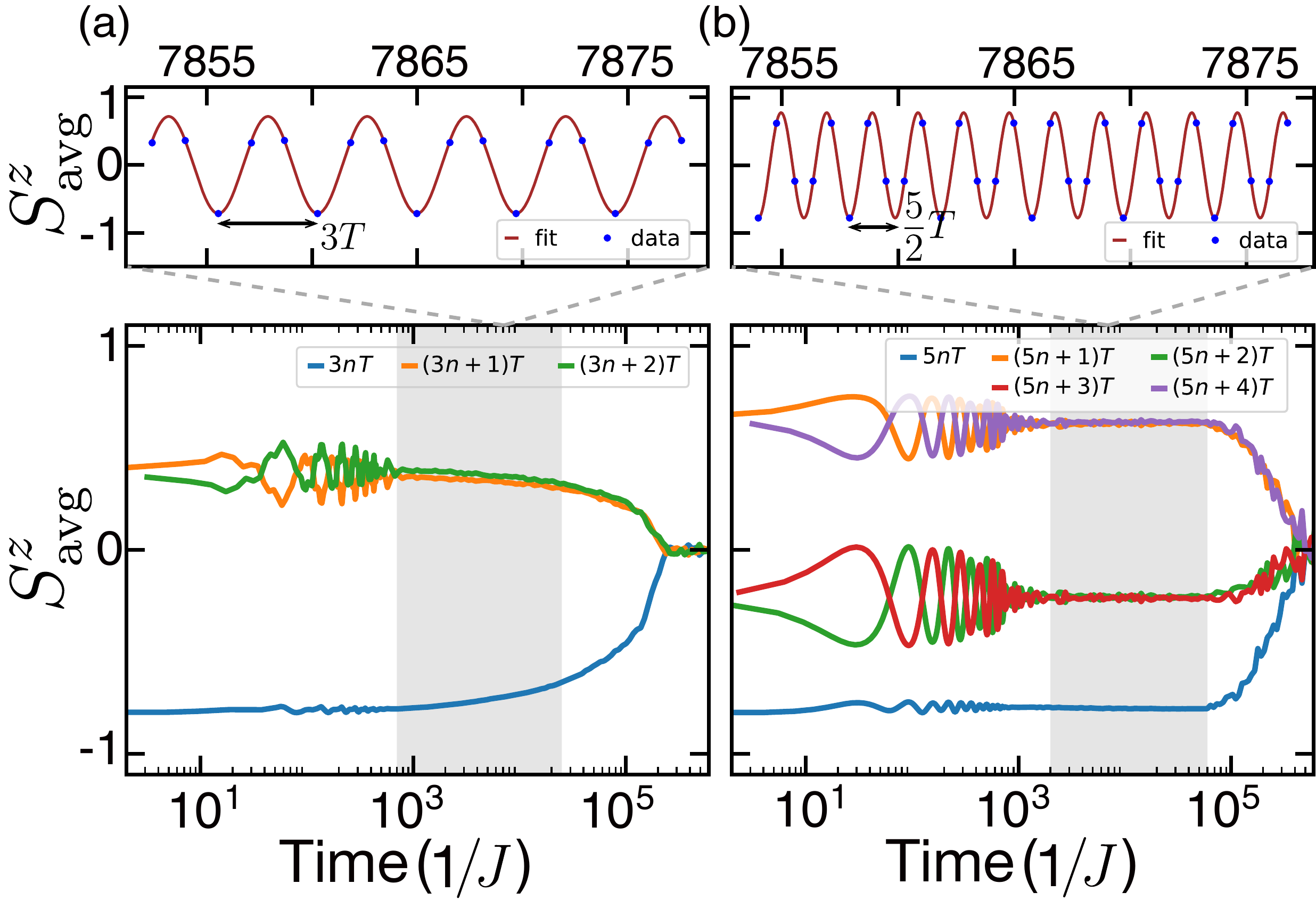}
    \centering
    \caption{Dynamics of higher-order CPDTC in a long-range interacting spin chain. 
    Lower panels: the magnetization exhibits an oscillation with a period of $3T$ [(a)] or $\frac{5}{2}T$ [(b)] in the prethermal regime (shaded regions) until the exponentially late the Floquet heating times. 
    Upper panels: zoom-in of magnetization over a few driving cycles in the prethermal regime. 
The parameters chosen in the simulations are $\{\alpha,J_z,J_{zz},J_x,J_y,h_x,h_y,h_z,gT,\omega\}=\{1.5,-0.383,0.53,0.28,0.225,0.13,0.09,0.06,2\pi/3,4.0\}$ and $\{\alpha,J_z,J_{zz},J_x,J_y,h_x,h_y,h_z,gT,\omega\}=\{1.5,-0.383,0.58,0.13,0.225,0.03,0.03,0.06,4\pi/5,8.0\}$ for the $3$-DTC and the $\frac{5}{2}$-DTC, respectively.
    }
    \label{fig:fraction_DTC}
  \end{figure}

Following exactly the same recipe for the 2-DTC, we now choose $M=3$ and thus the effective Hamiltonian will have an emergent $\mathbb{Z}_3$ symmetry. 
Initializing the system with a low-energy state (with respect to the effective Hamiltonian), we observe a local equilibration process followed by a subharmonic response with a period of $3T$, which can be diagnosed by the magnetization, Fig.~\ref{fig:fraction_DTC}(a). 

When $M$ is chosen to be a fraction, the dynamics are slightly different from the integer case. 
Previous work studied a similar situation in a nearly all-to-all interacting system, and observed signatures of robust subharmonic response at fractional frequencies \cite{pizzi2021higher}. 
These results can be framed within the CPDTC construction presented: namely, by choosing $M=5/2$, the effective Hamiltonian satisfies a $\mathbb{Z}_5$ symmetry, however, the rotation under $H_X(t)$ performs a $4\pi$-rotation every $5$ driving cycles instead of a single $2\pi$-rotation as considered in the previous examples.
Crucially, this results is a fractional time translational symmetry breaking with a new period of $5T/2$ (Fig.~\ref{fig:fraction_DTC}b). 
This is in stark contrast to the case when $M=5$: while in both cases, the energy shell of the effective Hamiltonian breaks into $5$ disjoint regions, the symmetry operation brings the system between these regions in different orders. 
As a result, the observation of a different Fourier spectrum for each of two phases relied on the fact that the magnetization is treated as a continuous variable which imbues the visited symmetry sectors with a particular order.

\bibliography{references}